\documentclass[12pt]{article}
\usepackage{a4wide}
\usepackage{graphics}
\usepackage{graphicx}
\usepackage{epsfig}
\usepackage{amssymb}
\usepackage{amsthm}
\usepackage{lineno}
\usepackage{amsmath}
\usepackage{color}
\usepackage[colorlinks=true, urlcolor=navyblue, linkcolor=navyblue, citecolor=navyblue]{hyperref}

\newcommand{\Xp}{\Xi_{cc}^+}
\newcommand{\Xpp}{\Xi_{cc}^{++}}
\newcommand{\Lc}{\Lambda_c^+}

\newcommand{\eV}{\mbox{\,eV}}
\newcommand{\keV}{\mbox{\,keV}}
\newcommand{\MeV}{\mbox{\,MeV}}
\newcommand{\GeV}{\mbox{\,GeV}}
\newcommand{\TeV}{\mbox{\,TeV}}

\newcommand{\fb}{\mbox{\,fb}}
\newcommand{\fs}{\mbox{\,fs}}
\newcommand{\fm}{\mbox{\,fm}}

\definecolor{navyblue}{rgb}{0,0.08,0.45}

\begin{document}

\thispagestyle{empty} 

\begin{flushright}
SLAC-PUB-17156
\end{flushright}

\begin{center} 
{\bf\large Resolving the SELEX--LHCb Double-Charm Baryon Conflict:\\[-5pt]
The Impact of Intrinsic Heavy-Quark Hadroproduction \\
and Supersymmetric Light-Front Holographic QCD}\\[1.0cm]
{\large S.J.~Brodsky$^1$, S.~Groote$^2$ and S.~Koshkarev$^2$}\\[0.3cm]
$^1$ SLAC National Accelerator Laboratory, Stanford University,\\
  Stanford, California 94309, USA\\[3pt]
$^2$ Institute of Physics, University of Tartu, 51010 Tartu, Estonia
\end{center}

\vspace{0.2cm}
\begin{abstract}
In this paper we show that the intrinsic heavy-quark QCD mechanism for the
hadroproduction of heavy hadrons at large $x_F$ can resolve the apparent
conflict between measurements of double-charm baryons by the SELEX
fixed-target experiment and the LHCb experiment at the LHC collider. We show
that in fact both experiments are compatible, and that both can be correct.
The observed spectroscopy of double-charm hadrons is in agreement with the
predictions of supersymmetric light front holographic QCD.
\end{abstract}
\newpage

\section{Introduction}

The first experimental evidence for the existence of double-charm baryons was
published by the SELEX collaboration 15 years ago~\cite{Mattson:2002vu,%
MattsonPhD,Moinester:2002uw,Ocherashvili:2004hi,Engelfried:2005kd,%
Engelfried:2007at}. By utilizing the Fermilab negative and positive charged
beams at $600\GeV/c$ to produce charmed particles in a thin foil of copper or
on a diamond target, the SELEX collaboration observed two different decay
channels for the $|dcc\rangle$ state at a mass close to $3520\MeV/c^2$. 

The SELEX fixed-target experiment measured hadron production in the forward
kinematic domain $x_F>0.1$. The negative beam composition was about 50\%
$\Sigma^-$ and 50\% $\pi^-$, whereas the positive beams were composed of 50\%
protons. The experimental data recorded used both positive and negative beams:
67\% of the events were induced by $\Sigma^-$, 13\% by $\pi^-$, and 18\% by
protons. In the first observation using the sample of
$\Lc\to pK^-\pi^+$~\cite{Kushnirenko:2000ed,Kushnirenko} SELEX found a signal
of $15.9$ events over $6.1\pm 0.1$ background events in the channel 
$\Xp\to\Lc K^-\pi^+$~\cite{Mattson:2002vu}. 
To complement this result, SELEX published an observation of $5.62$ signal
events over $1.38\pm 0.13$ background events for the decay mode $\Xp\to pDK^-$
from a sample of $D^+\to K^-\pi^+\pi^+$ decays~\cite{Ocherashvili:2004hi}.  

Two charm quarks will not be produced at high $x_F$ from DGLAP
evolution~\cite{Gribov:1972ri,Dokshitzer:1977sg,Altarelli:1977zs} or
perturbative gluon splitting
$g\to g+g\to(\bar cc)+(\bar cc)$~\cite{Altarelli:1977zs,Curci:1980uw,%
Furmanski:1980cm}. Therefore, the observation by SELEX of a double-charm
baryon $|qcc\rangle$ at a large mean value for $x_F$ and a relatively small
mean transverse momentum can raise skepticism. However, the $\Lambda_c(udc)$
and the $\Lambda_b(udb)$ were both discovered at the ISR at high
$x_F$~\cite{Lockman:1979aj,Chauvat:1987kb,Bari:1991ty}. In addition, the NA3
experiment measured both single-quarkonium hadroproduction
$\pi A\to J/\psi X$~\cite{Badier:1983dg} and double-quarkonium hadroproduction
$\pi A\to J/\psi J/\psi X$~\cite{Badier:1982ae} at high $x_F$. In fact, all of
the $\pi A\to J/\psi J/\psi X$ events were observed by NA3 with total
$x_F>0.4$. 

The existence of heavy quarks at large light-front (LF) momentum fraction $x$
in the proton's light-front wavefunction is in fact predicted by QCD if one
analyzes the higher Fock states $|uud c\bar c\rangle$ and
$|uud c\bar cc\bar c\rangle$ in the hadronic eigenstate; i.e., Fock states
where the heavy quark pairs are multi-connected to the valence quarks. LF
wavefunctions, the eigensolutions of the QCD LF Hamiltonian, are defined at
fixed LF time $\tau=t+z/c$, and are thus off-shell in invariant mass. For
example, in QED, positronium has an analogous $|e^+e^-\mu^+\mu^-\rangle$ Fock
state due to the insertion of light-by-light scattering in the positronium
self-energy amplitude. In such an ``intrinsic charm" Fock state
$|uudc\bar c\rangle$, the maximum kinematic configuration occurs at minimum
invariant mass, where all quarks are at rest in the hadron's rest frame; i.e.,
at equal rapidity in the moving hadron. Equal rapidity implies
$x_i\propto(m^2+{\vec k_\perp}^2)^{1/2}$ for each quark, so that the heavy
quarks in the Fock state carry most of the hadron's LF momentum. The operator
product expansion predicts that the probability of intrinsic heavy-quark Fock
states $|uud Q\bar Q\rangle$ to scale as $1/m_Q^2$ due to the non-Abelian
couplings of QCD~\cite{Brodsky:1984nx,Franz:2000ee}.

If such a Fock state interacts in a hadronic collision, the comoving $udc$ of
the projectile proton can readily coalesce into the $\Lambda_c(udc)$ bound
state with $x_F^\Lambda=x_c+x_u+x_d>0.4$. Similarly, it is natural to
hadro-produce a double-charm baryon $|qcc\rangle$ at high $x_F$ from the
materialization of the double-intrinsic charm $|uudc\bar cc\bar c\rangle$
Fock state of the projectile proton, since the $qcc$ quarks can coalesce at
the same rapidity~\cite{Brodsky:2012zza}. The production rate for double-charm
baryons is thus related to double quarkonium hadroproduction as observed by
NA3, since they both originate from the same double-intrinsic charm Fock state
of the projectile.

Recently, the LHCb collaboration published an observation of $313\pm 33$
events of $\Xpp\to\Lc K^-\pi^+\pi^+$ in a $13\TeV$ sample at the LHC and
$113\pm 21$ events in a $8\TeV$ sample at mass $3621.40\pm 0.72\text{(stat)}
\pm 0.27\text{(sys)}\pm 0.14(\Lc)\MeV/c^2$, corresponding to $1.7\fb^{-1}$ and
$2\fb^{-1}$, respectively~\cite{Aaij:2017ueg}. LHCb reported that the mass
difference between the  $\Xp(dcc)$ candidate reported by SELEX and the
$\Xpp(ucc)$ state reported by LHCb was $103\MeV/c^2$, so these states
cannot be readily interpreted as an isospin doublet since one would expect a
mass difference of isospin partners to differ by only a few $\MeV/c^2$. Note,
though, that the upper limit of the $x_F$ range at the LHCb collider
experiment is given by $x_F\approx 0.15$ and $x_F\approx 0.09$ for the $8\GeV$
and $13\GeV$ analysis, respectively (see discussion in Sec.~5). In contrast to
this, the $x_F$ range at the SELEX fixed-target experiment starts at $x_F=0.1$,
nearly complementary to the acceptance for the LHCb. In this paper we review
the hadroproduction mechanisms of double-charm baryons for the different
experimental environments and reinterpret the SELEX and LHCb results.


\section{Production rate and the kinematics of the $\Xp$\\
  for the SELEX experiment}

The SELEX collaboration did not provide the absolute production rate for the
double charmed baryon state $|dcc\rangle$. Fortunately, this rate can be
compared to that of $\Lc$ baryon. The production ratio $R_{\Lc}$ measured by
SELEX is given by
\[
R_{\Lc}^{\text{SELEX}} = \frac{\sigma(\Xp) \cdot
  Br(\Xp \to \Lc K^- \pi^+) }{\sigma(\Lc)}
  =\frac{N_{\Xp}}{\epsilon_+} \cdot \frac{\epsilon_{\Lc}}{N_{\Lc}},
\]
where $N$ is the number of events in the respective sample, and the
reconstruction efficiency of $\Xp$ is given by
$\epsilon_+\approx 11\%$~\cite{Mattson:2002vu}. The central value for the
number $N_{\Lc}/\epsilon_{\Lc}$ of reconstructed $\Lc$ baryon events reported
in Ref.~\cite{Garcia:2001xj} lies between $13326$ and $10010$ according to
whether the lowest bin with $x_F\in [0.125,0.175]$ is taken into account or
not. Therefore, we obtain
\[
R_{\Lc}^{\rm SELEX}\approx 0.012-0.014.
\]
If we take into account the intrinsic charm mechanism, the reconstruction
efficiency of $\Xp$ will grow at least by a factor of $2.3$ mainly because the
$x_F$ distribution predicted by intrinsic charm at large Feynman $x_F$ is well
matched to the acceptance of the SELEX fixed-target
experiment~\cite{MattsonPhD} (cf.\ Ref.~\cite{Groote:2017szb}). As a
consequence, $R_{\Lc}^{\rm SELEX}$ can be even smaller,
$R_{\Lc}^{\rm SELEX}\sim(0.5-0.6)\times 10^{-3}$.

By using the pion beam at $150$ and $280\GeV/c$ at CERN
incident on hydrogen and platinum targets, the NA3 experiment provided data on
double $J/\psi$ production with very similar features: a value for the ratio
$\sigma(\psi \psi)/\sigma( \psi)=(3\pm 1)\times 10^{-4}$ one or two orders of
magnitude higher than the conventional prediction of perturbative
QCD~\cite{Kiselev:2001fw}, values $x_{\psi \psi}>0.6$ at $150\GeV/c$ and
$x_{\psi \psi}>0.4$ at $280\GeV/c$ for the total Feynman-$x_F$ and total
transverse momentum $p_{T,\psi\psi}=0.9\pm 0.1\GeV/c$ of the
$J/\psi$ pair~\cite{Badier:1982ae,Badier:1985ri}.

Note that gluon--gluon fusion, quark--antiquark annihilation and the
gluon splitting mechanism of perturbative QCD can explain neither the NA3
cross section nor the $x_F$ distribution (see Ref.~\cite{Koshkarev:2017txl}
and references therein). Both double $J/\psi$ and doubly-charmed baryons could
be produced perturbatively by mechanisms such as double parton scattering or
some kind of ladder diagram where e.g.\ gluons emitted from the projectile and
the target both split into $Q\bar Q$ pairs, one heavy quark from each pair
being connected by an intermediate gluon. This is of course not the same as
double parton scattering but leads to the same outcome in terms of producing
two $Q\bar Q$ pairs in the same hard scattering. However, these pairs would
typically have lower average energies, softer $p_T$ distributions and narrower
$x_F$ ranges, particularly at fixed-target energies. They are also reduced in
yield by additional factors of the strong coupling constant. However, intrinsic
charm has none of these deficiencies but just a lower probability for the
production of the second $Q\bar Q$ pair, while it still produces relatively
large $x_F$ heavy flavor quarks.

It is clearly of interest to relate the production of the $\Xp$ at the
SELEX experiment with the production of double $J/\psi$ production at the NA3
experiment. Unfortunately, it is not possible to compare the two results
directly. However, we are able to compare the following ratios
$R=\sigma(c\bar cc\bar c)/\sigma (c\bar c)$:
\[
R^{\rm SELEX}=R_{\Lc}\times\frac{f(c\to\Lc)}{f_{\Xi_{cc}}}
  \sim(1-4)\times 10^{-3}
\]
\vspace{-7pt}\noindent
and
\vspace{-7pt}
\[
R^{\rm NA3}=\frac{\sigma(\psi\psi)}{\sigma(\psi)}\times
  \frac{f_\psi}{f^2_{\psi/\pi}}\sim 2\times 10^{-2},
\]
where $f_{\psi/\pi}\approx 0.03$ is the fragmentation rate of the intrinsic
charm state of the pion into $J/\psi$~\cite{Vogt:1995tf} and
$f_\psi\approx 0.06$ is the perturbative QCD fragmentation rate into
$J/\psi$~\cite{Mangano:2004wq}.
$f_{\Xi_{cc}}\approx 0.25$~\cite{Koshkarev:2016rci} represents the fraction of
double $c\bar c$ pairs producing the sum of single-charged baryons $\Xp$ and
double-charged baryons $\Xpp$, but this fraction cannot be less than the
fraction to produce $J/\psi$. Therefore, $R^{\rm SELEX}$ should not be larger
than $10^{-2}$. The SELEX production ratio is thus in approximate consistency
with the complementary measurement of the double $J/\psi$ production by the
NA3 experiment. It is interesting to note that the intrinsic charm mechanism
predicts $\langle x_F(\Xi_{cc})\rangle=0.33$, as shown in
Ref.~\cite{Koshkarev:2016rci}. This is in excellent agreement with the value
$\langle x_F(\Xp)\rangle\sim 0.33$ measured by the SELEX experiment.

It should be emphasized that SELEX observed the weak decay of the $3520\MeV$
double-charm baryon in two different decay channels, namely
$\Xp(3519\pm 1)\to\Lambda_c^+K^-\pi^+$ and $\Xp(3518\pm 3)\to pD^+\pi^-$ with
statistical significances of 6.3 $\sigma$ and 4.8 $\sigma$, respectively. The
probability that these two signals are statistical fluctuations is extremely
small.

\section{Mass difference}
In order to resolve the discrepancy between the results from SELEX and LHCb
we will utilize the predictions of the supersymmetric light front holographic
QCD (SUSY LFHQCD). This approach was developed by imposing the constraints
from the superconformal algebraic structure on LFHQCD for massless
quarks~\cite{Dosch:2015nwa}. As has been shown in
Refs.~\cite{Dosch:2015nwa,Dosch:2015bca}, supersymmetry holds to a good
approximation, even if conformal symmetry is strongly broken by the heavy
quark mass.

Note that the $3_C+\bar 3_C$ diquark structure of the $\Xp$ can be written
explicitly as $|[dc]c\rangle$ state~\cite{Koshkarev:2018kre}. The production
of the double-charm baryon $\Xi_{[dc]c}^+$ with $[dc]$ in a spin-singlet state
is natural in the SELEX fixed target experiment since it has acceptance at
high $x_F$, i.e.,\ in the realm of intrinsic charm; the $[dc]c$ configuration
can easily re-coalesce from a higher Fock state of the proton such as
$|uudc\bar cc\bar c\rangle$. In contrast, the production of this state is
likely to be suppressed in $q\bar q\to c\bar cc\bar c$ or
$gg\to c\bar cc\bar c$ reactions at the LHCb. Thus LHCb has most likely
observed the double-charm baryon state $|u(cc)\rangle$, as will be explained
in the next section.\footnote{We use square brackets $[\ ]$ for spin-0 and
round brackets $(\ )$ for spin-1 internal states.} The mass difference between
the $|[dc]c\rangle$ and the $|u(cc)\rangle$ states is due to the hyperfine
interaction between the quarks.

Supersymmetric light front holographic QCD, if extended to the case of two
heavy quarks, predicts that the mass of the spin-1/2 baryon should be the same
as the mass of $h_c(1P)(3525)$ meson~\cite{Dosch:2015bca}. This is well
compatible with the SELEX measurement of $3520.2\pm 0.7\MeV/c^2$ for the
$\Xp(d[cc])$, although the uncertainty of SUSY LFHQCD predictions is at least
of the order of $100\MeV$. Indeed, the mass of the $|u(cc)\rangle$ state is
predicted to be the same as that of the $\chi_{c2}(1P)(3556)$ meson, which is
in turn lower than the LHCb result of
$3621.40\pm 0.72\text{(stat)}\pm 0.27\text{(sys)}\pm 0.14(\Lc)\MeV/c^2$
for the $\Xpp$.

Supersymmetric LFHQCD is based on and best tested in the chiral limit of QCD,
where all quarks are massless. The mass difference between the $h_c(1P)$ and
the $\chi_{c2}(1P)$ is mainly due to the hyperfine splitting between the two
charm quarks, and hence very small. In the baryon there is also the larger
spin-spin interaction between the $c$ and a light quark. By comparing hadron
masses with light and charmed quarks, one can estimate the strength of this
additional, the supersymmetry-breaking interaction in the range
$84-136\MeV/c^2$~\cite{Dosch:2015bca, Brodsky:2016yod}, which is well
compatible with the mass difference between the SELEX and the LHCb states.

\section{The SELEX state at the LHCb}
In the previous section we identified the SELEX state as a $|[dc]c\rangle$
state and the LHCb state as a $|u(cc)\rangle$ state of the double charmed
baryon. While the SELEX state is definitely a spin-1/2 state, both $J^P=1/2^+$
and $J^P=3/2^+$ are possible assignments for the LHCb state. As becomes clear
in the following, $J^P=1/2$ is favored by the LHCb mass measurement and the
very suppressed radiative decay to $|[qc]c\rangle+\gamma$. Based on the Heavy
Quark Effective Theory (HQET), in Ref.~\cite{Korner:1994nh} the baryon masses
are estimated to be $3610\MeV/c^2$ for the spin-1/2 state
$|u(cc)\rangle_{1/2^+}$ and $3680\MeV/c^2$ for the spin-3/2 state
$|u(cc)\rangle_{3/2^+}$. For a qualitative estimate one can also compare with
the nucleon states where the lowest mass $I=1/2$, $J^P=3/2^+$ state is the
$N(1720)$ which is considerably more massive than the proton.

It is interesting to analyze the ability of the LHCb experiment to observe
the $|dcc\rangle$ state, i.e.\ $\Xp$. Note that in a sample corresponding to
$0.65\fb^{-1}$ of integrated luminosity at $7\TeV$, in case of the decay
process $\Xp\to\Lc K^-\pi^+$ the LHCb collaboration published an upper limit
for the ratio $\sigma(\Xp)\cdot Br(\Xp\to\Lc K^-\pi^+)/\sigma(\Lc)$ of
$1.5\times 10^{-2}$ and $3.9\times 10^{-4}$ for the lifetimes $100\fs$ and
$400\fs$, respectively~\cite{Aaij:2013voa}. In case of the decay
$\Xpp\to\Lc K^-\pi^+\pi^+$ analyzed in Ref.~\cite{Aaij:2017ueg}, one expects a
larger lifetime, $\tau(\Xpp)/\tau(\Xp)\approx 2.5 - 4$ (cf.\
Refs.~\cite{Kiselev:2001fw,Karliner:2014gca}). Taking into account that the
cuts were optimized for the lifetime of $333\fs$ and that the minimum lifetime
reached by the LHCb is about three times larger than the lifetime
$\tau(\Xp)<33\fs$ measured by the SELEX at 90\% confidence level, the LHCb
provided an analysis which was outside the signal region for $\Xp$.

It is also of interest to analyze the expectation of the production ratio
between states. As we discussed above, the production of the double-charm
baryons at the LHCb is due to $q\bar q\to c\bar cc\bar c$ or
$gg\to c\bar cc\bar c$ reactions with the following fragmentation of the
$cc$-diquark into the double-charm baryons. Due to the Pauli principle the
$cc$-diquark has to be a state with spin 1 or higher, leading to a state
$|q(cc)\rangle$ with higher mass. The normalization of the fragmentation of
the $cc$-diquark into the double-charm baryons is unknown. However, we are
still able to provide some quantitative analysis. The fragmentation function
is proportional to the wave function at the origin. The color-anti-triplet
wave function can be estimated on the basis of information about the
color-singlet wave function,
$|R(0)[cc]_{\bf\bar 3}|\sim|R(0)[c\bar c]_{\bf 1}|$. It is clear that the
$|[qc]c\rangle$ states cannot be produced through the fragmentation mechanism
on the LHCb.

\section{Suppression of the radiative decay}
An important issue is the rate for the heavier $|u(cc)\rangle$ state to decay
radiatively to the spin-1/2 ground state $|[uc]c\rangle_{1/2^+}$. However,
LHCb explicitly observed that the state they discovered decays weakly which
means that the radiative decay of the LHCb state at $3621\MeV$ has to be
strongly suppressed: The radiative lifetime has to be longer than approximately
$50\fs$ in order that at least some of the LHCb $3621\MeV$ states would have
survived and observed to decay weakly. A lifetime of $50\fs$ means that the
transition rate $\Gamma(3621\to 3520+\gamma)$ has to be less than $0.01\eV$.
The energy of the photon emitted by the radiative transition is
$\omega=101\MeV$. The dependence of the transition rate on $\omega$ comes from
(a) the phase space of the final state and (b) the dynamical suppression from
the square of the matrix element of the electromagnetic current connecting the
initial and final eigenstates.

For comparison, one can consider the measured radiative decay rate for
$J/\psi\to\eta_c\gamma$ with photon energy
$\omega=115\MeV$~\cite{Ebert:2002pp}. The measured radiative decay rate for
$J/\psi\to\eta_c\gamma$ is $\Gamma^{\rm exp}=1.13\pm 0.35\keV$. Note that the
spatial wavefunctions of the $J/\psi$ and the $\eta_c$ are almost identical.
In contrast, the initial and final state wavefunctions are very different. In
the case of transition between the double-charm baryons
$|q(cc)\rangle_{3/2^+}\to|[qc]c\rangle_{1/2^+}+\gamma$ the amplitude for
radiative decay thus involves the matrix element
$\langle f|\vec J_{\rm em}|i\rangle$ of the electromagnetic current between
highly orthogonal hadronic eigenstates. In particular, the emission of the
$\omega=101\MeV$ photon also has to interchange one of the charm quarks in the
spin-1 $(cc)$ diquark with the light quark $q$ to form the spin-0 $[qc]$
diquark.     

The matrix element in the LF framework involves the overlap of the current
with nearly orthogonal three-body light-front wavefunctions
$\psi_i(x,\vec k_\perp)$ and $\psi_f(x,\vec k_\perp)$ at the small $101\MeV$
momentum transfer. The matrix element thus must vanish as $(\omega r_{cc})^3$,
where $r_{cc}$ characterizes the size of the radial wavefunctions. This gives
a strong suppression of the rate of $(\omega r_{cc})^6$ relative to that of
radiative charmonium decays.

According to Ref.~\cite{Nochi:2016wqg},
$r_{J/\psi}\sim 0.39\fm\simeq 2\GeV^{-1}$. As a first estimate, we will
assume that the characteristic radial size of the charm diquark $cc$ in the
double-charm baryons is the same as that of the $J/\psi$,
$r_{cc}\sim r_{J/\psi}$, resulting in $(\omega r_{cc})^6\sim 6\times 10^{-5}$.
In comparison with the $1\keV$ $J/\psi\to\eta_c\gamma$ decay width, this would
give a radiative transition rate for
$|q(cc)\rangle_{3/2^+}\to|[qc]c\rangle_{1/2^+}+\gamma$ of order
$0.06\eV$ and thus a radiative transition lifetime of $10\fs$. A radiative
lifetime of this order would not prevent the LHCb from observing the weak
decay of the $|q(cc)\rangle$ double-charm baryon at $3621\MeV$. In this case
LHCb might be able to observe some radiative events
$|[qc]c\rangle_{1/2^+}+\gamma$, where the $|[qc]c\rangle_{1/2^+}$ at
$3520\MeV$ decays weakly. 

It is also possible that the $3621\MeV$ state observed by LHCb is a
$J^P=1/2^+$ double-charm baryon state $|q(cc)\rangle_{1/2^+}$, rather than
$J^P=3/2^+$ since we have assumed that it is a bound state of a spin-1/2 quark
and a spin-1 $(cc)$ diquark. In this case the spin-1 photon needs to be
emitted with orbital angular momentum $L=1$ to conserve parity in the
radiative decay of the $3621\MeV$ state $|q(cc)\rangle_{1/2^+}$ to the
$3520\MeV$ state $|[qc]c\rangle_{1/2^+}$. This would give an additional factor
of $v^2$ in the phase space for the radiative transition rate where
$v\simeq\omega/M\sim 1/35$ is the recoil velocity of the $3520\MeV$
double-charm baryon. This additional suppression of the rate implies that the
lifetime of the radiative transition would then be increased to be of order
$12000\fs$.

We have also done a comparison with the radiative transition rates between
the $J^P = 1^-$ charmonium $S$ states analyzed in the classic paper by
Feinberg and Sucher~\cite{Feinberg:1975hk}, such as the estimated $10\keV$
transition rate between the $\psi(2S)(3686\MeV)$ and the $J/\psi(3097\MeV)$.
The transition energy in our case is $101\MeV$, compared to the transition
energy between the $\psi(2S)(3686\MeV)$ and the $J/\psi(3097\MeV)$ of
$589\MeV$ -- a relative reduction in the photon transition energy of
$101/589\sim 1/6$. This factor enters at the third power in the matrix element
of the spin current in Eq.~(9) of Ref.~\cite{Feinberg:1975hk} through the
overlap of radial wavefunctions, i.e.\ the spherical Bessel function
$j_0(kr/2)$. In addition, one has to take into account the additional
suppression of electromagnetic transitions  between the $q(cc)$ and $[qc]c$
configurations when extrapolating to double-charm baryons.

The above discussion is clearly only a first estimate. A rigorous treatment of
the transition radiative decay rate between double-charm baryons with diquarks
of different spin and composition is clearly necessary.  

We also note that LHCb may be be able to detect radiative transitions
involving double-charm baryons which have higher masses and higher spin.
The observation of the $3520\MeV$ state $|[uc]c\rangle_{1/2^+}$ with a
significant transverse momentum kick from photon emission from a heavy double
charm state would be a important confirmation of our picture. Because of this,
we suggest that LHCb conduct such a search.

Still, there might be intrinsic charm in the wave function at LHC. To analyze 
this opportunity we will follow a similar discussion given in
Ref.~\cite{Koshkarev:2016rci}. The contribution from the double intrinsic
charm should be suppressed due to the kinematics of the LHCb experiment:
Making the naive assumption that the momentum is split evenly between all
final states and taking into account that the hadron identification efficiency
for pions and kaons is degraded above
$100\GeV/c$~\cite{Adinolfi:2012qfa,Papanestis:2017zcj}, the analysis loses
sensitivity around $p(\Xpp)\approx 600\GeV/c$, i.e.\ $x_F\approx 0.15$ and
$x_F\approx 0.09$ for the $8\GeV$ and $13\GeV$ analysis, respectively. This
range of values for $x_F$ corresponds to the rapidity region $2<y<5$ in which
the LHCb detector operates~\cite{Alves:2008zz}. Note that in contrast to
SELEX, LHCb is a collider experiment where the acceptance excludes the
detection of events close to the beam axis. 

\section{Summary}
Using both theoretical and experimental arguments, we have shown that the
SELEX and the LHCb results for the production of doubly charmed baryons can
both be correct.  We have compared the data for double $J/\psi$ production
observed by the NA3 experiment and the SELEX result for $\Xp$ production at
high Feynman-$x_F$. We have found that the NA3 data strongly complement the
SELEX production rate for the spin-1/2 $|[dc]c\rangle$ state. In contrast,
LHCb has most likely discovered the the heavier $|u(cc)\rangle$  
produced by gluon--gluon fusion $gg\to c\bar cc \bar c$ at $x_F\sim 0$. The
application of supersymmetric algebra to hadron spectroscopy, together with
the intrinsic heavy-quark QCD mechanism for the hadroproduction of heavy
hadrons at large $x_F$, can thus resolve the apparent conflict between
measurements of double-charm baryons by the SELEX fixed-target experiment and
the LHCb experiment at the LHC collider. The mass difference of the two
double-charm baryons reflects the distinct spins of the underlying diquarks.

An important conclusion from our study is that the natural kinematic domain
for producing novel hadronic bound states, such as multi-heavy quark hadrons
and the tetraquarks predicted by superconformal algebra, is large Feynman
$x_F$. In this domain, the constituents of the higher Fock states
of the projectile, which are comoving at the same rapidity, can coalesce to
produce a wide variety of color-singlet hadrons.

Our paper shows why the state $|[qc]c\rangle_{1/2^+}$ is favorably produced at
high $x_F$, within the kinematics of the SELEX acceptance, and conversely, why
its production is unfavorable in the LHC acceptance. There are also very
strong upper limits for the production of double-charm baryons in forward
photoproduction from duality which explains why the FOCUS experiment does not
observe any signal for double charm baryons~\cite{Ratti:2003ez}.

These observations indicate the importance of the high $x_F$ domain and
fixed-target LHC experiments such as SMOG@LHCb~\cite{Aaij:2011er,Aaij:2014jba}
and AFTER@LHC~\cite{Brodsky:2012vg,Koshkarev:2016acq} for observing the
hadroproduction of exotic heavy quark states. As explained above, one also
can understand why the radiative transition
$3621|(cc)q\rangle\to 3520|[qc]c\rangle+\gamma$ is strongly suppressed due to
the very small overlap of their respective radial wavefunctions.

The SELEX experiment measured the lifetime $\tau(\Xp) < 33\fs$ at 90\%
confidence level. Using the ratio $\tau(\Xpp)/\tau(\Xp)\approx 2.5 - 4$ we can
assume $\tau(\Xpp) \approx 100\fs$. This value stands in contrast to
the minimum values theoretically predicted as $\tau(\Xp)\approx 53\fs$ and
$\tau(\Xpp)\approx 185\fs$~\cite{Karliner:2014gca}. Therefore, definitive
measurements of the lifetime will provide another test for the SELEX data.

\subsection*{Acknowledgements}
This research  was supported by the Estonian Research Council under Grant
No.~IUT2-27 and by the Department of Energy under Contract
No.~DE--AC02--76SF00515. The authors would like to thank H.G.~Dosch and
G.F.~de T\'eramond for critical comments.

\end{document}